\documentclass[prl,aps,twocolumn,showpacs]{revtex4}
\input epsf
\usepackage{graphics}
\usepackage{graphicx}

\begin{document}
\title{Nonlocal Measurements in the Time-Symmetric Quantum Mechanics.}

\author{ Lev Vaidman and Izhar Nevo}


\affiliation
{ School of Physics and Astronomy\\
Raymond and Beverly Sackler Faculty of Exact Sciences\\
Tel-Aviv University, Tel-Aviv 69978, Israel}

\date{}

\vspace{.4cm}

\begin{abstract}
  Although nondemolition, reliable, and instantaneous quantum
  measurements of some nonlocal variables are impossible, demolition
  reliable instantaneous measurements are possible for all variables.
  It is shown that this is correct also in the framework of the
  time-symmetric quantum formalism, i.e. nonlocal variables of
  composite quantum systems with quantum states evolving both forward
  and backward in time are measurable in a demolition way. The result
  follows from the possibility to reverse with certainty the time
  direction of a backward evolving quantum state. Demolition
  measurements of nonlocal backward evolving quantum states require
  remarkably small resources. This is so because the combined
  operation of time reversal and teleportation of a local backward
  evolving quantum state requires only a single quantum channel and no
  transmission of classical information.
\end{abstract}


\maketitle

Causality imposes  constraints on nondemolition quantum
 measurements \cite{PV}. It  allows measuring only certain
 classes of nonlocal variables \cite{AAV86}.
Recently, it has been shown that  it is possible to measure
instantaneously an arbitrary nonlocal
variable of two spin$-{1\over 2}$ particles \cite{GR} and
of any composite system \cite{V03}, provided we do not require nondemolition 
measurement.  Such demolition measurement
distinguishes with certainty between eigenstates of the nonlocal
variable, but, contrary to the standard  quantum
measurement, it might destroy the measured eigenstate. The
ability of performing such measurement provides physical meaning
for nonlocal variables in the framework of relativistic quantum
mechanics.

 Recently, we  witness  significant developments of
the  time-symmetric formalism of quantum mechanics \cite{AV90,AV02}
originated by Aharonov, Bergmann and Lebowitz \cite{ABL}. This
formalism contains, in addition to the standard quantum state evolving
forward in time, the quantum state evolving
backward in time from complete measurement performed in the future
relative to the time in question. Moreover, in this framework one
can define states evolving for one subsystem in one time direction
and in the opposite time direction for another. Thus, one can define
novel types of nonlocal variables, for which the eigenstates evolve
backward in time or even in both time directions.  In this work
we show that such nonlocal variables are also measurable (in the
sense of demolition measurements) and this result provides physical
meaning for nonlocal variables and states of the time-symmetric
quantum formalism.

Let us now state more precisely the problem we want to consider. The
quantum system consists of two or more parts separated in space. We
assume that we are equipped with measuring devices which allow 
measuring any local variable in each part of the system. We also have
unlimited resources of entanglement, the quantum channels which
connect the sites in which separate parts of the system are located.
Given  unlimited time, these resources would allow, via
teleportation of the quantum states of the different parts to one location, a
measurement of an arbitrary variable. However, we require {\em
  instantaneous} measurement and thus there is no time to complete the
teleportation of the quantum states of the different parts. The meaning of
``instantaneous" is that shortly after the time of measurement, there
are permanent records at the locations of the various parts of the system,
which jointly
 yield the value of the nonlocal variable.
 Examples of nonlocal observables are:  
variables with entangled eigenstates such as Bell operator, 
variables with product eigenstates which cannot be measured via
measurements of local variables (which got the name {\em
nonlocality without entanglement}) \cite{NLWE}, etc.
 Some nonlocal
variables are measurable in a stronger sense of standard
nondemolition quantum measurements, but 
 there are nonlocal
variables which are not measurable in this sense \cite{AAV86}. This is why we
consider demolition
measurements which can be performed on all variables.

Let us recall the basic idea of nonlocal demolition measurements
\cite{V03} in the case of a bipartite system where each part consists
of several spin$-{1\over 2}$ particles.  It consists of a sequence of
``half teleportations'' between the sites of the system, where, say,
Alice and Bob are located. ``Half teleportation" consists of the Bell
measurement performed on the system and one particle from the EPR pair
(the quantum channel). It does not include the transmission of the
result of the Bell measurement and the appropriate correction of the
state of the second particle of the EPR pair. Thus,
``half-teleportation'' does not have a minimal time for
implementation.

 We note that if a quantum state has a simple product
form:
 \begin{equation}\label{PrSt}
 |\Psi\rangle =\Pi_i|s_i\rangle_i ,
\end{equation}
where $|s_i\rangle$ is either $|{\uparrow}_z\rangle$ or
$|{\downarrow}_z\rangle$, then it can be unambiguously determined in a
local measurement, even if it was first ``half teleported" from
another place. The information required for this includes the results
of the spin measurements of the ``half-teleported'' state together with
the results of the Bell measurements of the ``half-teleportation''
procedure. Indeed, the only possible effects of the second half of
teleportation procedure are changes of the total phase and spin flips.
The phase is irrelevant, and the effect of the spin flip can be
corrected later, when the results of the Bell measurements arrive.
 
The procedure goes as follows. At the beginning, the state of Bob's
part of the system is ``half-teleported" to Alice. Alice assumes that
the ``half-teleportation" is, in fact, the full teleportation (there
is a finite probability that the results of Bell measurements
correspond to teleportation without correction) and transforms the
eigenstates of the nonlocal variable to the orthogonal set of states
in the form (\ref{PrSt}). Then, she ``half-teleports'' the transformed
state to Bob. Bob knows the results of his teleportation Bell
measurements and, if those were successful, he performs the spin
measurements in the $z$ basis. Together with the results of Alice
teleportation Bell measurements these results yield the eigenstate of
the measured variable. If Bob's first teleportation was not
successful, he teleports the received state back to Alice in one of
the numerous quantum channels according to the results of the Bell
measurements in the first teleportation attempt. Now, Alice assumes
that Bob's second teleportation was successful and unitarily
transforms the outputs of all Bob's channels to the form (\ref{PrSt})
and then teleports each transformed state in separate channel to Bob.
Every round adds finite probability for success, and the probability
of success after many rounds converges to 1. The method can be
generalized to any number of sites of a composite quantum system.

The method cannot be applied for measurements of backward evolving
states by simple ``time-reversal" of our operations. Indeed, in our
procedure, Bob, starting from the second round, teleports the
quantum state in a particular channel depending on the results of
his previous Bell measurement. In the time reversed procedure, this
corresponds to choosing the channel according to results of a
measurement which has not been performed yet. Note, that standard
von Neumann measurement {\em is} applicable for measurement of
backward evolving quantum state. Here, the same procedure measures quantum
states evolving forward and backward in time. The same holds  also for
nondemolition nonlocal quantum measurements \cite{AAV86}. Unfortunately, such
measurements can be performed only for a very limited class of
nonlocal variables. The question remains: Is it possible to
measure (in a demolition way) an arbitrary variable for a backward
evolving quantum state?

The answer is positive, and, moreover, we can measure in a
demolition way any variable also for quantum states evolving in
different directions at different locations. The solution is very
simple: the quantum state evolving backward in time can be
transformed to a state evolving forward in time. For
quantum states evolving forward in time there is a measurement
procedure which leads to a successful measurement with any desired
probability. Therefore, there is  a method for
measuring  a nonlocal variable for quantum state with parts
evolving in arbitrary time directions.

  Any quantum system with a finite number of states can be mapped onto
a system of $N$ spin$-{1\over 2}$ particles. So, all what we have to 
do is to change the time direction of a spin$-{1\over 2}$ particle.
  To
this end, we perform a Bell measurement on the particle and an
ancilla and then (if needed) local operations acting on the ancilla so as
to change the state of the two particles to a singlet ${1\over \sqrt
2}( |{\uparrow}\rangle |{\downarrow}\rangle
-|{\downarrow}\rangle|{\uparrow}\rangle)$. In fact, we can prepare
the singlet in any other way too. We also have to ensure that there
is no disturbance between the time of the Bell measurement and the time
we receive the particle with the backward evolving state. Then, the
ancilla obtains the time reversed state of the particle:
\begin{equation}\label{TR}
   \alpha \langle{\uparrow} |+\beta\langle{\downarrow}| \rightarrow
   -\beta^\ast |{\uparrow}\rangle +\alpha ^\ast
|{\downarrow}\rangle .
\end{equation}
When we perform such time reversal for all parts of the system
evolving backward in time,  the eigenstates of the nonlocal variable
with parts of the system evolving backward in time are transformed
to a well defined mutually orthogonal states evolving forward in
time. Thus, we obtain a one to one correspondence between the
eigenstates of the variable with parts evolving in different time
directions and the eigenstates of a variable evolving forward in time.
The latter we know how to measure in a demolition way, thus, we got
a  method for a demolition  measurement of an arbitrary
nonlocal variable.
\begin{figure}[b]
\label{fig1}
\includegraphics[totalheight=5cm,width=7.5cm]{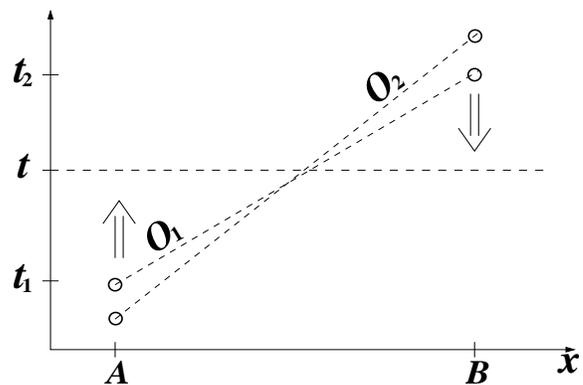}
\caption{The ``preparation'' of quantum state evolving forward in time in $A$ and backward in time in $B$.}
\end{figure}


Let us describe an example. In order to ``prepare" an eigenstate of
an operator corresponding to a quantum state evolving in part $A$
forward in time and in part $B$ backward in time consider two
``crossed" nonlocal (in space and time) measurements, Fig.1:

\break

\begin{eqnarray}
\label{Prepa}
 \nonumber O_1 &\equiv & \sigma_z^{A}(t_1) - \sigma_z^{B}(t_2), \\
O_2 &\equiv &(\sigma_x^{A}(t_1-\epsilon) -
\sigma_x^{B}(t_2+\epsilon)){\rm mod} 4.
\end{eqnarray}
\noindent
These measurements yield for time $t,~~t_1<t<t_2$, one of the
eigenstates of a nonlocal variable evolving forward in time in part
$A$ and backward in time in part $B$:

\begin{eqnarray}\label{ES}
\nonumber O_1&=&2:~~~~~~~~~~~~~~~ |{\uparrow}\rangle_A~~ \langle{\downarrow}|_B ,\\
   O_1 &=&-2:~~~~~~~~~~~~~ |{\downarrow}\rangle_A~~ \langle{\uparrow}|_B ,\\
\nonumber  O_1&=&0,~~O_2=0:~~~ {1\over \sqrt 2} (
|{\uparrow}\rangle_A~~ \langle{\uparrow}|_B +
   |{\downarrow}\rangle_A~~ \langle{\downarrow}|_B),\\
 \nonumber O_1&=&0,~~O_2=2:~~~ {1\over \sqrt 2}( |{\uparrow}\rangle_A~~
 \langle{\uparrow}|_B -
   |{\downarrow}\rangle_A~~ \langle{\downarrow}|_B).
\end{eqnarray}
\noindent
Now, our time reversal operation (\ref{TR}) at time $t-\epsilon$ at part $B$
transforms these eigenstates to
\begin{eqnarray}\label{TES}
\nonumber &~&|{\uparrow}\rangle_A~~ |{\uparrow}\rangle_B ,\\
  \nonumber &~& |{\downarrow}\rangle_A~~ |{\downarrow}\rangle_B ,\\
  &~& {1\over \sqrt 2} ( |{\uparrow}\rangle_A~~ |{\downarrow}\rangle_B
  -
   |{\downarrow}\rangle_A~~ |{\uparrow}\rangle_B) ,\\
 \nonumber  &~& {1\over \sqrt 2} ( |{\uparrow}\rangle_A~~ |{\downarrow}\rangle_B
  +
   |{\downarrow}\rangle_A~~ |{\uparrow}\rangle_B)) .
\end{eqnarray}
These are the eigenstates of  a nonlocal variable evolving forward in time which cannot be
measured in a nondemolition way \cite{PV}, but can be measured in
a demolition way.

Although the method explained above allows to  measure
any nonlocal variable, it requires unnecessary large resources.
In order to make a reliable measurement of a forward evolving quantum
state with parts located in several places, we need a huge amount of
entanglement \cite{V03}. Amusingly, if in a particular location the
quantum state evolves backward in time, the measurement 
can be performed in a simpler way, and  a lot of entanglement resources can be saved.
Our task is bringing the states of all parts to one
location. The quantum state evolving backward in time can be moved
to another place in one step which also includes the change of the
direction of the time evolution of the state. All what we have to do
is to prepare the EPR pair with our particle and an ancilla located in
the place to which  we want to move the state of the particle. When
there is a large distance between the two places this might require a
significant time, but we have this time since we can prepare the EPR pair in
advance. If we cannot arrange in advance that one particle from the 
EPR pair is part of our post-selected system, we can
always perform local swap operation between the member of the EPR
pair and the post-selected particle. Thus, the backward evolving
quantum state or, the state with only one part evolving forward in
time, can be measured in a demolition way using very moderate
resources: $N-1$ quantum channels for $N$-part system.

Until now we discussed quantum states evolving at each space-time
point in a single time direction. In the framework of the time symmetric
formalism of quantum mechanics \cite{AV90}, the complete description
of a quantum system is given by a two-state vector $\langle
\Phi|~~|\Psi\rangle$ such that there are both forward and backward
evolving quantum states at each part of the system. The most general
description of a composite quantum system of $N$ parts is the {\em
generalized two state vector} \cite{AV91}:
 \begin{equation}\label{GTSV}
\sum_i \alpha_{(i_1,..i_N,j_1,..j_N)} \prod_{n=1}^N \langle
\Phi_{i_n}|_n~~|\Psi_{j_n}\rangle_n .
\end{equation}
A generalized two-state vector might not be reducible to a two-state
vector $\langle \Phi|~~|\Psi\rangle$. In this case, in order to
``prepare" such a state at time $t$, it is necessary to have  ancilla
particles protected from the time of the the joint measurement of our
system and the ancilla performed before $t$ and until  the joint
post-selection measurement performed after time $t$. (Sufficiently large 
system    can always be  described  by a two-state vector,
like in standard quantum mechanics, sufficiently large 
system  can always be described by a pure quantum state.)

The $N$-part system described by a generalized two-state vector is
equivalent to $2N$-part system with quantum state in half of the
parts evolving forward in time and in half of the parts evolving
backward in time. Thus, preparing singlets with ancilla particles in
the location of one of the parts of the system and their partners at
the locations of the other parts, we can bring all backward evolving
parts of the quantum state to our chosen location. From this point
we proceed with the procedure of measuring nonlocal forward
evolving state of $N$-part system \cite{V03}.

The conceptual possibility of measuring an arbitrary variable with
eigenstates which are nonlocal generalized two state-vectors
provides justification of ascribing physical meaning to nonlocal
two-state vectors. Since all nonlocal variables can be
measured, the projection operators on all two-state
vectors are measurable.

 We now  discuss further  the concepts described above.
  What do we exactly mean by a ``backward evolving
state"? We need not assume that at present there is a state actually 
 evolving backward in time from a future measurement, i.e., that the
future is fixed.  The formalism is helpful even in a pragmatic
approach when we consider a scenario with post-selection: At a
particular time we perform measurements on each member of an
ensemble of quantum systems. Later, somebody performs another
measurement and {\em post-selects} a subensemble of systems with a
particular result of his measurement. There are no constrains on the
post-selection measurement: any local or nonlocal variable can be
measured, since there is no requirement that it will be
instantaneous. All parts of the system can be brought to one place
and then, by assumption, any measurement can be performed. Then, we
discard all the results for systems which were not post-selected and
consider only systems which were successfully post-selected.

A basic requirement for a measurement, and in particular, for  our
demolition nonlocal measurement, is that it reliably distinguishes
the eigenstates of an observed variable, i.e., our measurement and
the post-selection should yield the same results. Then, it seems
that a measurement of a variable for a backward evolving quantum state
can be performed in a much simpler way: just {\em prepare} an
eigenstate of this variable. Then, the post-selected ensemble will
consists only of the cases in which the post-selection measurement
yields the same result. Preparation of nonlocal state requires much
less resources than its verification, since we can prepare the state
of all parts of the system locally in one place and then bring them
to the separate locations at the right time. (We assume through out the paper
that the free Hamiltonian is zero. In practice we can always
compensate for the free evolution.)

However, the preparation of a particular eigenstate is not a good enough
verification measurement of a backward evolving state. What we ask
from a quantum measurement is not only that it yields the correct
value of the measured variable if the state is the eigenstate of
this variable. Rather, we  also need that if the state is a superposition of
different eigenstates, then the measurement yields one of them with
the appropriate probability. Thus, if the backward evolving state is
a superposition of several eigenstates of the measured variable,
then the preparation of one particular state yields incorrect probability.

Backward evolving state defines probability of a measurement at
present only if there is no quantum state evolving forward in time
from the measurements in the past. This is usually not the case
(while it is a usual situation for the time-reversed situation in
which forward evolving quantum state is considered since future is
considered to be unknown). Thus the past of the quantum system has
to be erased, it should be unknown. This can be done using
collective measurements on the system and ancilla particles. After erasing the past we can prepare at random (using another ancilla) eigestates of the  measured variable such that the probability will end up to be correct, but
then, this procedure is probably not simpler than  the one described
above.

What allowed us to solve the problem of measurability of nonlocal
variables with arbitrary directions of time evolution for various
parts of the system is a simple, but apparently new result: {\em we
can change the time direction with certainty from backward evolving
quantum state to forward evolving quantum state}.

Note that this is not so if we try to transform forward to backward
evolving state. In order to reverse the forward evolving state we
need to obtain singlet as a result of Bell measurement performed on
our particle and an ancilla. The probability for this is $1\over 4$.
If the result is different, we cannot correct it to a singlet, since
this operation depends on the results of Bell measurement, but has
to be performed {\em before} the Bell measurement.

Although we do not consider here continuous systems (every real
system can be approximated well as a system with a finite number of
states) it is interesting to note that we can, conceptually,
perform a time reversal of a backward evolving quantum state of a
continuous system $\Psi (q)$ by preparing the original EPR state (the state of
the Einstein, Podolsky, and Rosen paper \cite{EPR}) for the particle and
an ancilla:
\begin{equation}\label{EPRst}
   |q-\tilde{q}=0, ~~p+\tilde{p}=0\rangle .
\end{equation}
Then, the backward evolving quantum state of the particle will
transform into a complex conjugate state of the ancilla:
\begin{equation}
\label{CTR}
   \Psi(q) \rightarrow \Psi^\ast(\tilde{q}).
\end{equation}
If the particle and the ancilla are located in different locations,
then such operation is a combination of time reversal and
teleportation of a backward evolving quantum state of a continuous
variable \cite{V94}.

It is a pleasure to thank  Shmuel Nussinov for helpful discussions. This research was supported in part by grant 62/01 of the Israel
Science Foundation.



\end{document}